# Hybrid Integration of Quantum Dot Single-Photon Sources with Lithium Tantalate Photonics for On-Chip Routing

Kaili Xiong[1†], Defeng Shan[2†], Xueshi Li[1], Ziliang Ruan[2], Bin Chen[2], Zhanling Wang[3], Jiawei Wang[4], Ying Yu[5], Wei Wu[1], Pingxing Chen[1], Jin Liu[5*], Liu Liu[2*], Yan Chen[1*], Tian Jiang[1,6*]

[1]Institute for Quantum Science and Technology, National University of Defense Technology, Changsha, 410073, China.
[2] College of Optical Science and Engineering & State Key Laboratory of Extreme Photonics and Instrumentation, Zhejiang University, Hangzhou 310058, China
[3]College of Electronic Science and Technology, National University of Defense Technology, Changsha, 410073, China.
[4]School of Integrated Circuits, Harbin Institute of Technology (Shenzhen), Shenzhen 518055, China.
[5]State Key Laboratory of Optoelectronic Materials and Technologies, Sun Yat-sen University, Guangzhou, 510275, China.
[6]Hunan Research Center of the Basic Discipline for Physical States, Changsha, 410073, China.

[†]These authors contribute equally to this research.
[*]Corresponding authors.
Emails: liujin23@mail.sysu.edu.cn; liuliuopt@zju.edu.cn; chenyan@nudt.edu.cn; tjiang@nudt.edu.cn

**Abstract**

A promising pathway towards scalable quantum photonic processors involves the simultaneous integration of deterministic single-photon sources, low-loss photonic circuitry, and fast reconfigurability. Thin-film lithium tantalate on insulator (LTOI) offers an exceptional electro-optic response and low optical loss at 900 nm wavelength band, yet its lack of efficient quantum emitters has hindered progress toward fully integrated quantum technologies. Here, we demonstrate heterogeneous integration of indium arsenide quantum dots (QDs) with low-loss reconfigurable LTOI waveguides (0.30±0.04 dB/cm) using micro-transfer printing. By directly butt-coupling tapered gallium arsenide waveguides with inversely tapered LTOI waveguides, we achieve robust and alignment-tolerant inter-waveguide coupling. The hybrid chip operates at cryogenic temperatures, enabling deterministic routing of successively emitted single photons from the QDs with a half-wave voltage-length product (~1.9 V·cm at 4 K), confirming the cryogenic stability of LTOI's electro-optic coefficient. These results establish the first demonstration of high-speed on-chip routing of single photons with hybrid QD-LTOI circuits, providing a scalable pathway toward integrated quantum photonic processors.

## Introduction

Ferroelectric lithium compounds, such as lithium niobate (LN) and lithium tantalate (LT), are cornerstone materials in photonics due to their outstanding electro-optic, piezoelectric, and nonlinear properties. Recent advances in thin-film lithium niobate on insulator (LNOI) have enabled high-quality-factor resonators (*1–4*) and broadband modulators on chip (*5, 6*). Notably, the successful integration of cryogenic-compatible LNOI circuits with superconducting nanowire single-photon detectors (SNSPDs) has underscored their immense potential for quantum technologies (*7–10*). Sharing the same crystal structure with LNOI, thin-film lithium tantalate on insulator (LTOI) also demonstrates remarkable modulation and nonlinear optical properties. Moreover, ferroelectric thin-film platforms like LT offer exceptional properties for integrated quantum photonics, including stable electro-optic response at cryogenic temperatures and low optical propagation losses at the emission wavelength of the quantum emitters (~900 nm) (*11–14*). These characteristics are crucial for building low-loss and reconfigurable photonic integrated circuits that operate effectively with on-chip quantum emitters. Various integrated photonic devices with enhanced performance have been demonstrated based on LTOI, such as electro-optic modulators with ultralow drift (*15*) and ultrabroadband electro-optic frequency comb (*16*). Despite these milestones, a fundamental limitation persists: LNOI or LTOI platforms inherently lack efficient quantum emitters. While rare-earth ions can be doped into these ferroelectric lithium compounds, isolating single ions remains challenging due to the low emission rates of intra-4f transitions (*17, 18*). Consequently, developing on-chip deterministic light sources presents a significant hurdle.

In contrast, III-V semiconductor quantum dots (QDs) offer compelling advantages as single-photon sources (*19–22*). State-of-the-art QD devices achieve high photon extraction efficiencies (*23*) and near-unity photon indistinguishability (*24, 25*), positioning them as ideal candidates for integrated quantum photonic processors. Nevertheless, most research focuses on standalone devices. Interfacing these with photonic circuits typically requires fiber coupling to external components for quantum photonic operations (*26–29*). This off-chip approach incurs substantial photon loss, severely degrading the already weak single-photon flux. On-chip integration is thus essential for scalable photonic quantum information processing (*30*).

Heterogeneous integration strategies offer highly viable solutions (*31–35*). For instance, QDs have been integrated with passive platforms like silicon nitride (SiN) waveguides (*36, 37*). However, the lack of active tunability in SiN fundamentally limits its utility in reconfigurable quantum circuits, particularly at cryogenic temperatures. While significant progress has been made in hybrid integration, demonstrated by Aghaeimeibodi et al. with LNOI confirming preserved single-photon properties (*38*), and Hugo et al. with foundry silicon photonics enabling scale (*39*), these demonstrations have primarily focused on proof-of-concept coupling and passive transmission. Crucially, no previous work has functionally harnessed the integrated circuit's inherent processing capabilities, such as high-speed electro-optic switching, to perform deterministic quantum operations on the single photons emitted on-chip. Consequently, the fundamental viability of using deterministic solid-state quantum emitters within a fully programmable, monolithically integrated photonic circuit remains an open question in the field.

Here, we bridge this critical gap by demonstrating the heterogeneous integration of III-V QDs with LTOI circuits. Using a high-precision micro-transfer printing technique, we implement a tapered end-to-end waveguide architecture that ensures robust, high-efficiency coupling between the QD emitters and the LTOI circuit. We show that single photons emitted from the QDs are directly coupled into the LTOI waveguides at cryogenic temperatures. Leveraging LT's cryo-electro-optic properties, we demonstrate high-speed single-photon routing with a half-wave voltage-length product of 1.9 V·cm at 4 K and low propagation loss (0.3 dB/cm @ 900 nm). This work represents a crucial step toward scalable, on-chip quantum information processing by providing a platform that combines a high-performance single-photon source with ultrafast, cryogenic-compatible electro-optic control.

## Results
### Device sketch and design

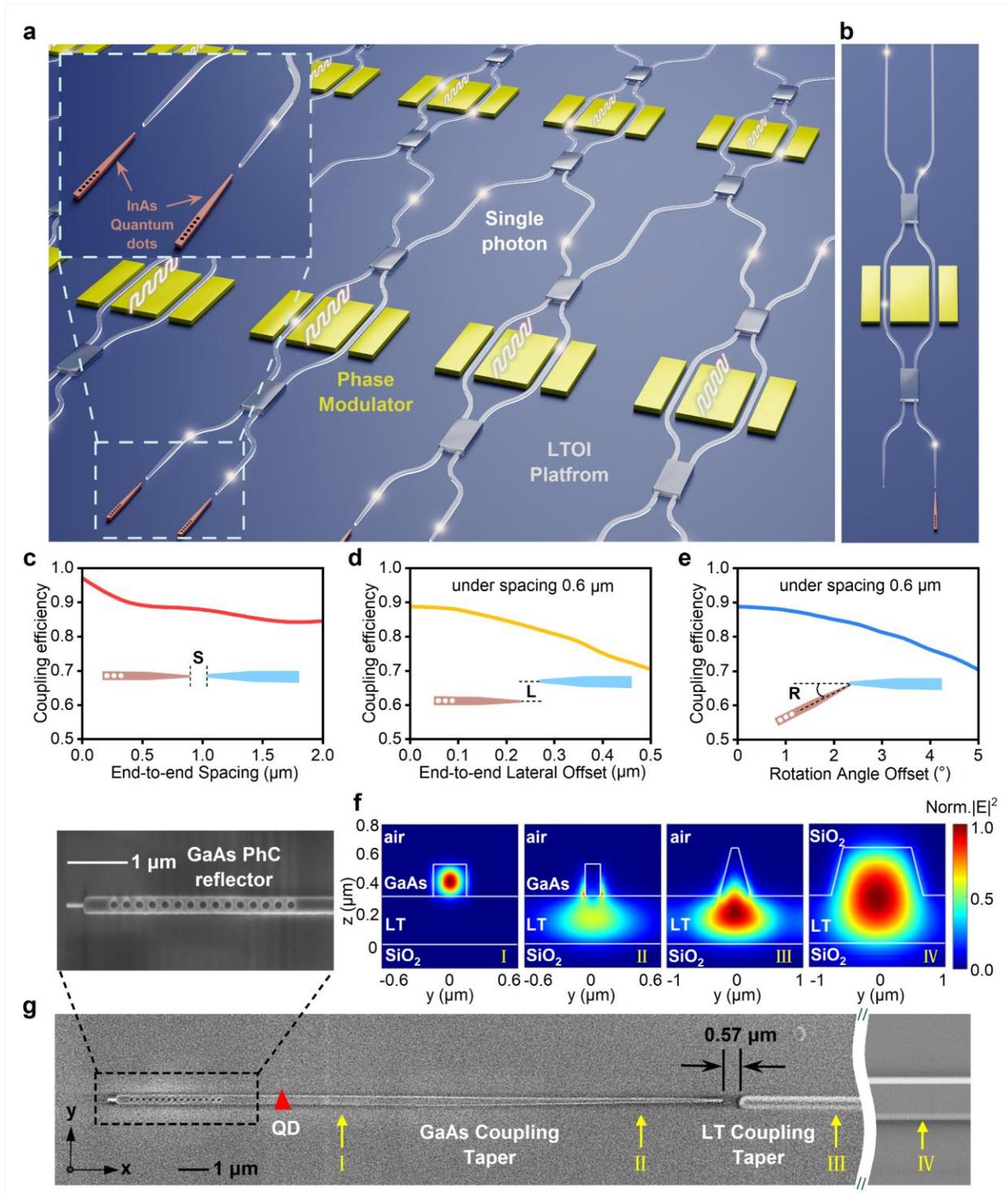

**Figure 1. Hybrid quantum photonic device architecture.** (a) Artistic schematic of the heterogeneous integration between indium arsenide (InAs) QDs and LT Mach-Zehnder interferometer (MZI). The inset shows a zoomed-in view of the butt-coupled tapered gallium arsenide (GaAs) and LT waveguides. (b) Schematic of the device configuration investigated in this work, comprising a single GaAs waveguide coupled to a single LT MZI. (c-e) Simulated coupling efficiency as a function of (c) end-to-end waveguide spacing, (d) lateral offset, and (e) relative angular misalignment. (f) Simulated cross-sectional electric field distributions at sequential planes (I to IV), illustrating the adiabatic mode transfer from the GaAs waveguide to the LTOI waveguide. Cross-section IV indicates the untapered LT waveguide. (g) False-colored scanning electron micrograph of the fabricated hybrid device: GaAs waveguide is aligned to the LT waveguide with sub-100 nm precision. The inset shows a photonic crystal reflector at the end of the GaAs waveguide.

The hybrid quantum photonic chip features an LTOI Mach-Zehnder interferometer (MZI) structure. The electro-optic response enables dynamic optical path manipulation for high-speed single-photon routing (*40*). The device operates at ~900 nm to match the emission wavelength of the indium arsenide (InAs) QD single-photon sources. Figure 1a illustrates the future architecture of this chip platform, where the spectral randomness of emitters can be addressed via local tuning mechanisms such as the Stark effect or strain tuning (*41*, *42*). Furthermore, the positional randomness of self-assembled QDs can be mitigated using grid-marker-assisted photoluminescence imaging combined with dual-color light-emitting diode (LED) co-illumination and electron-multiplying charge-coupled device (EMCCD) imaging, or via in situ cryogenic three-dimensional electron-beam lithography (*43*, *44*). These methodologies provide a robust pathway for the deterministic and scalable production of high-performance quantum photonic circuits.

In the present work, we investigate a single gallium arsenide (GaAs) waveguide coupled to a single LT MZI, as shown in Fig. 1b. To maximize the coupling efficiency between the QD and the LT waveguide, we implemented an in-plane butt-coupling scheme: tapered GaAs waveguides are aligned end-to-end with inversely tapered and shallowly etched LTOI waveguides. The GaAs tapers feature a 200-nm thickness and 20-μm length, with widths adiabatically tapering from 300 nm to 80 nm. A terminal GaAs photonic crystal mirror is placed at the other end to enhance the coupling efficiency. This planar design outperforms vertical evanescent coupling in misalignment tolerance, maintaining >80% efficiency at 2 μm separation (Fig. 1c). Coupling efficiency decreases from 88% to 71% as lateral displacement L increases from 0 to 0.5 μm (Fig. 1d). Besides, the coupling efficiency is also not sensitive to relative angle of the two waveguides, maintaining above 70% even for 5° (Fig. 1e). These results confirm robustness against typical fabrication misalignments (±0.1 μm tolerance), which is advantageous for scalable manufacturing.

Fig. 1f maps the simulated electric field distributions across sequential cross-sections. Light remains confined within the GaAs waveguide, gradually expanding through its adiabatic taper before encountering the inversely tapered LTOI waveguide. Mode profiles at each cross section (Fig. 1g) are plotted in Fig. 1f. The corresponding LTOI waveguide mode profile confirms geometric matching between the structures. This modal overlap ensures resonant cavity matching, enabling a high coupling efficiency. Fig. 1g shows a scanning electron micrograph of the hybrid device. The right waveguide denotes the LTOI waveguide, while the left is the transferred GaAs waveguide. The transfer process achieves high precision, with an inter-waveguide spacing of 0.57 μm and negligible lateral misalignment (<0.1 μm). A photonic crystal mirror fabricated at the end of the GaAs waveguide enhances photon collection efficiency (pattern shown in inset).

**The hybrid chip fabrication**

The modulator was fabricated on a commercial x-cut LTOI wafer comprising a 600 nm-thick LT layer, a 4.7 μm buried oxide layer, and a 525 μm undoped silicon substrate. Two multimode interferometers serving as 3 dB couplers for the MZM structure are symmetrically positioned at both ends of the L = 3 mm long travelling wave electrodes. Edge couplers are used to connect the chip to external near-infrared single-mode fibers. The ridge waveguides in the modulation region are 300 nm high and 1 μm wide. In the coupling region, the LTOI waveguide is gradually tapered to 100 nm with a length of 10 μm.

Fig. 2a shows the optical image of the fabricated chip with a 3 mm classic coplanar waveguide traveling-wave electrode on an undercut etched silicon substrate. The electrode with the ground-signal-ground configuration is adopted in our device to enable collinear propagation of the optical and radio frequency (RF) waves. The top-left panel of Fig. 2a shows the cross-sectional view of the modulator. The gap $g$ between the signal and ground electrodes is the major factor influencing the modulation efficiency $V_\pi \cdot L$. However, it should be noted that while reducing the gap $g$ enhances modulation efficiency, the propagation loss caused by metal absorption of optical waves will also increase. To mitigate these adverse effects and optimize device performance, the metal electrode gap is set to $g$ = 6 μm. The signal electrode has a width of $w$ = 10 μm, which is wide enough to ensure a low microwave loss. Considering the device performance and cost, the thickness of the electrodes is $h$ = 0.9 μm. To protect the LTOI waveguide while suppressing metal-induced reflections and reducing optical propagation loss, a silicon dioxide over-cladding layer with thickness $f$ = 1.5 μm is deposited between the metal electrodes. Furthermore, the silicon substrate beneath the signal electrode and adjacent waveguides is removed via backside etching with a trench width of $r$ =17 μm, significantly reducing microwave loss by minimizing RF wave absorption in the substrate. Details on the modulator design and fabrication are described in Supplementary Note I.

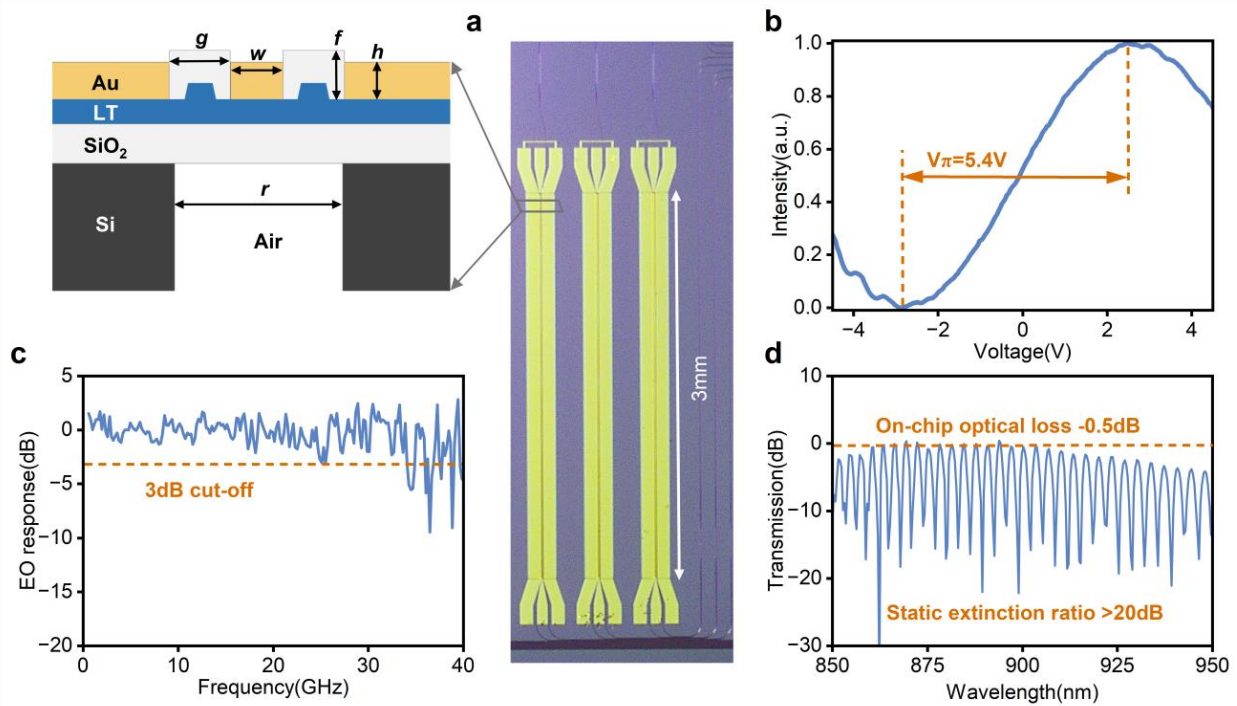

**Figure 2. Room-temperature electro-optic characterization.** (a) Schematic of the LTOI MZI with a coplanar waveguide travelling-wave electrode. Cross-sectional view of the MZI showing the key

parameters: *g* (electrode gap), *w* (signal electrode width), *h* (electrode thickness), *f* (oxide cladding thickness), and *r* (etching trench width). (b) Modulation curve showing the half-wave voltage ($V_\pi$) measurement. (c) Electro-optic frequency response showing a 3-dB bandwidth exceeding 40 GHz. (d) Measured optical transmission spectrum of the MZI device.

We fabricated freestanding GaAs waveguides on a 200 nm-thick GaAs membrane with embedded InAs QDs. The waveguides stand upon a 600 nm-thick sacrificial layer. A photonic crystal reflector is patterned at the end of each waveguide, which has a high reflection window between 900-950 nm. The reflector is used to enhance the coupling efficiency of light. After the fabrication of the waveguides, the sacrificial layer is wet-etched, leaving freestanding thin devices. Details on the method of GaAs waveguide fabrication are described in Supplementary Note II.

The waveguide is transfer-printed to the LTOI photonic chip using a translation stage. To enhance printing accuracy, we developed a micro-polydimethylsiloxane (PDMS) transfer printing method. By replacing conventional glass slides with thin coverslips during the transfer process, this modification enables real-time high-quality imaging of waveguides for precision alignment. Detailed methodology is provided in the Supplementary Note III.

**Photonic chip characterization at room temperature**

Comprehensive optical characterization of the integrated modulator was performed at room temperature. The chip was fiber-couped on the translation stage. Subsequent measurements employed a 915-nm laser source with transverse-electric (TE) polarization, demonstrating a half-wave voltage of $V_\pi$ = 5.4 V under 100-kHz triangular voltage scanning, corresponding to $V_\pi \cdot L$ = 1.62 V·cm (Fig. 2b). Small-signal electro-optic response characterization using a vector network analyzer and 40-GHz photodetector (Newport 1414) showed a 3 dB bandwidth exceeding 40 GHz (Fig. 2c), surpassing our detector's measurement range. RF modulation signals were applied through a probe to electrodes on one side, with the opposing side terminated by a 50 Ω load.

The total insertion loss of the device, including contributions from both input and output edge couplers, was measured to be 8.5 dB when injecting TE polarized light from a continuous-wave laser. To isolate the intrinsic on-chip performance, the fiber-to-chip coupling loss was quantified separately using an identical straight-waveguide reference structure fabricated on the same wafer, yielding a combined edge coupler loss of 8.0 dB. After subtracting this calibrated coupling loss, the net on-chip insertion loss of the MZI itself was determined to be 0.5 dB. This low value highlights the high quality of the fabricated LTOI waveguides and efficient multimode couplers. Furthermore, to assess the wavelength dependence across the QD emission window, the laser source was scanned from 850 nm to 950 nm while monitoring the transmitted optical power (Fig. 2d). This spectral scan provides a baseline for operating the device with broadband single-photon sources. Additionally, we characterized the passive propagation loss of the bare LT waveguides via the cut-back method, obtaining a value of ~0.3 dB/cm at 900 nm. Detailed procedures for the loss calibration are provided in Supplementary Note IV.

**On-chip photon routing at low temperatures**

Next, we characterized the device at cryogenic temperatures (4 K). A side-excitation with vertical-collection geometry was adopted for photoluminescence (PL) measurements, which is compatible with the device architecture. The sample was wire-bonded to an RF carrier instead of using electrical probes, with RF lines connected via cryostat feedthroughs. To determine the half-wave voltage of the modulator at cryogenic temperatures, above-band excitation of the QDs was performed using a continuous-wave laser. Various voltages were applied to the modulation electrode via a DC voltage source, and the photons from Output Port 1 were directed into a spectrometer for PL measurements. The resulting QD PL intensity exhibited periodic modulation with the applied voltage (Fig. 3a). Electro-optic characterization at cryogenic temperatures revealed a half-wave voltage $V_\pi$ = ~6.30 V (The corresponding $V_\pi \cdot L$ is ~ 1.89 V·cm). This value aligns with room-temperature measurements, indicating no degradation of LT's electro-optic coefficient at low temperatures.

To verify the routing capability at cryogenic temperatures, we modulated a continuous-wave (910 nm) laser at 100 MHz (slightly above our 80-MHz femtosecond excitation requirement). The collected signal detected by an avalanche photodiode (APD) exhibited sinusoidal modulation with ~90% depth, demonstrating maintained high-frequency response (Fig. 3b). The residual nonzero intensity primarily originates from bandwidth limitations in the cryogenic electrical packaging (e.g., impedance mismatch and parasitic inductance from wire bonds) rather than intrinsic device defects. A detailed frequency-dependent analysis of this modulation depth is provided in Supplementary Note VI.

For dynamic photon routing, we excited the QD using a 780-nm femtosecond laser operating at an 80-MHz repetition rate. The laser synchronization signal was delivered to an arbitrary function generator, which produced a rectangular wave switching between two distinct control voltages ($V_1$ and $V_2$) applied to the MZI electrodes (Fig.3c). This voltage switching enabled deterministic path selection for successive photons: the first emitted photon was routed to Output Port 1, while the subsequent photon was directed to Output Port 2. APD measurements at the output ports revealed pronounced anti-correlated counting events. Specifically, when monitoring Port 1, photon counts were exclusively detected during $V_1$ application periods (with negligible counts during $V_2$ phases). Conversely, Port 2 registered counts only during $V_2$ application (Fig.3d). This absence of cross-talk unambiguously confirms that successive photons were dynamically rerouted to distinct output paths at the laser repetition rate (80 MHz), experimentally validating the high-speed reconfigurability of our integrated quantum photonic circuit for deterministic single-photon routing.

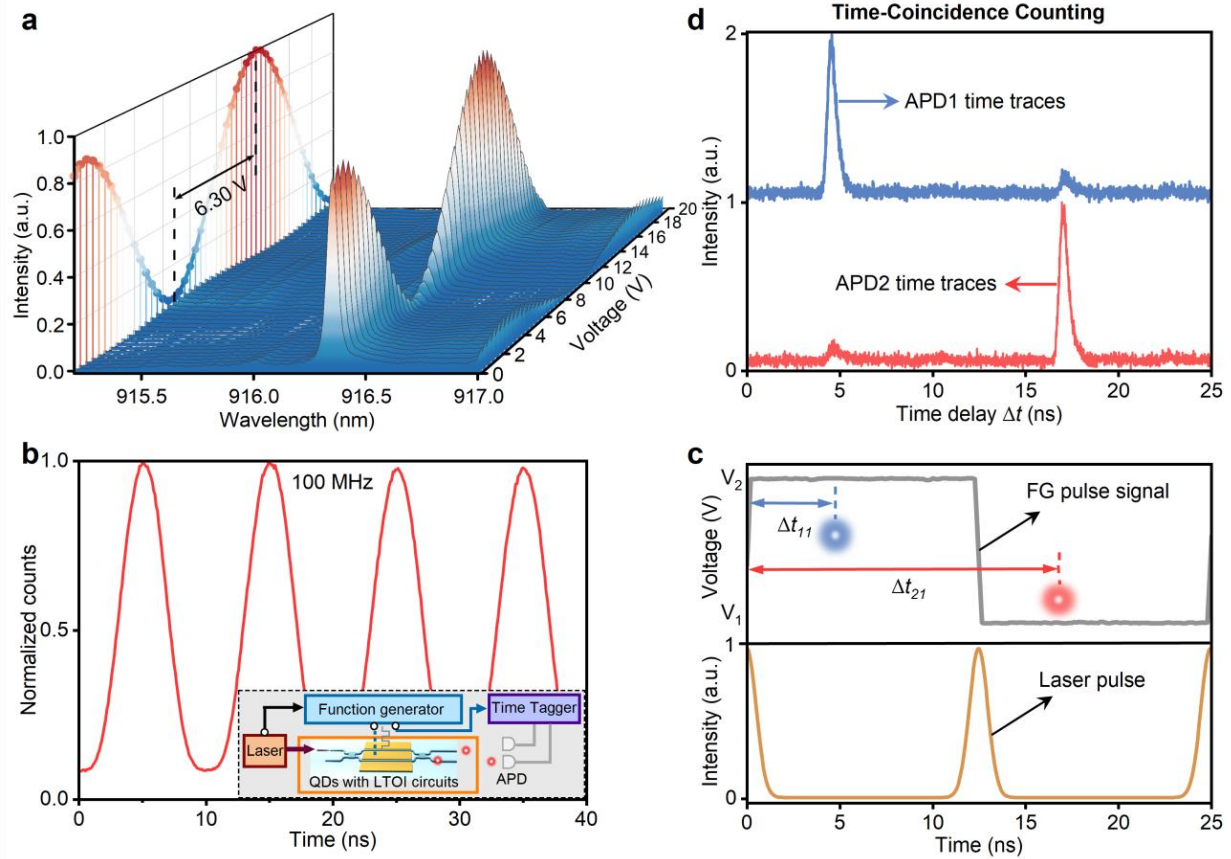

**Figure 3. On-chip single-photon routing at cryogenic temperatures.** (a) The spectra mapping of the QD's emission under different voltages applied to the modulation electrode by a DC voltage source. The left panel presents the cryogenic electro-optic characterization, revealing the measured half-wave voltage. (b) Example of a sinusoidal modulation of a continuous-wave laser at 100 MHz at 4 K, demonstrating the device's high-frequency response. The inset shows the experimental setup for on-chip single-photon routing. (c) The orange curve represents the pulsed laser output signal, while the grey curve indicates the rectangular wave signal synchronously output by the arbitrary function generator to the electrode. The blue and red photons denote their arrival times at the APD from different output ports, illustrating deterministic routing to distinct ports depending on the applied voltage. (d) Time traces showing the arrival of single photons at two separate output ports.

**Single photon characterization**

To examine the emission dynamics, we performed time-resolved lifetime measurements. The QD was excited using a 785 nm laser (repetition rate: 80 MHz), with the laser synchronization signal serving as the temporal reference for photon arrival time measurements. The measured lifetime decay curves are presented in Fig. 4a. The QD exhibited a radiative lifetime of $\tau = 930.0\pm0.1$ ps. This confirms the feasibility of GHz-rate single-photon routing in principle, though experimental demonstration is currently limited to 80 MHz by the repetition rate constraints of our laser source.

To quantify the efficiency of coupling single photons from the QD into the integrated photonic circuit, we performed power-dependent PL measurements (Fig. 4b). The collected PL intensity was calibrated against the laser excitation power while accounting for the known efficiency of the detection system. This analysis yielded an extraction efficiency at the first objective of 29.7%, representing the fraction of emitted photons successfully guided by the hybrid GaAs-LT waveguide system and collected by the objective lens. It is worth noting that this value constitutes a conservative estimation, as it does not account for the QD blinking effects - a known phenomenon reducing emitter brightness. And several fabrication-dependent factors also: (1) non-ideal reflectivity (<100%) of the GaAs photonic crystal mirror terminal, in contrast to the perfect mirror assumption in simulations; and (2) Sidewall roughness in the etched LT waveguides, which introduces additional scattering losses not captured in idealized models. These imperfections represent addressable challenges for future process optimization. The detailed calibration methodology, including correction factors for objective NA, filter transmission, and detector quantum efficiency, are discussed in implementation specifics (Supplementary Note VII).

We measured photon autocorrelation using a Hanbury Brown–Twiss (HBT) interferometer. Photons were directed to a beam splitter with outputs monitored by two single-photon detectors. Figure 4c plots the coincidence histogram, revealing strong antibunching at zero delay ($g^{(2)}(0) = 0.080 \pm 0.003$ after fitting), indicating suppressed multiphoton emission. This confirms the single-photon emission characteristics of the QD source. The residual $g^{(2)}(0)$ value possibly originates from imperfect laser rejection and QD re-excitation.

We also characterized the indistinguishability of the emitted single photons via a standard Hong-Ou-Mandel (HOM) two-photon interference experiment. The photons were sent into an unbalanced MZI with a path-length difference matching the excitation laser repetition period. As shown in Fig. 4(d), we recorded the coincidence counts for consecutive photons prepared in completely distinguishable (cross-polarized) and indistinguishable (co-polarized) states. The raw two-photon interference visibility is determined to be $V = 1 - A_\parallel/A_\perp = 76.9 \pm 0.4\%$ (detailed in Supplementary Note VIII).

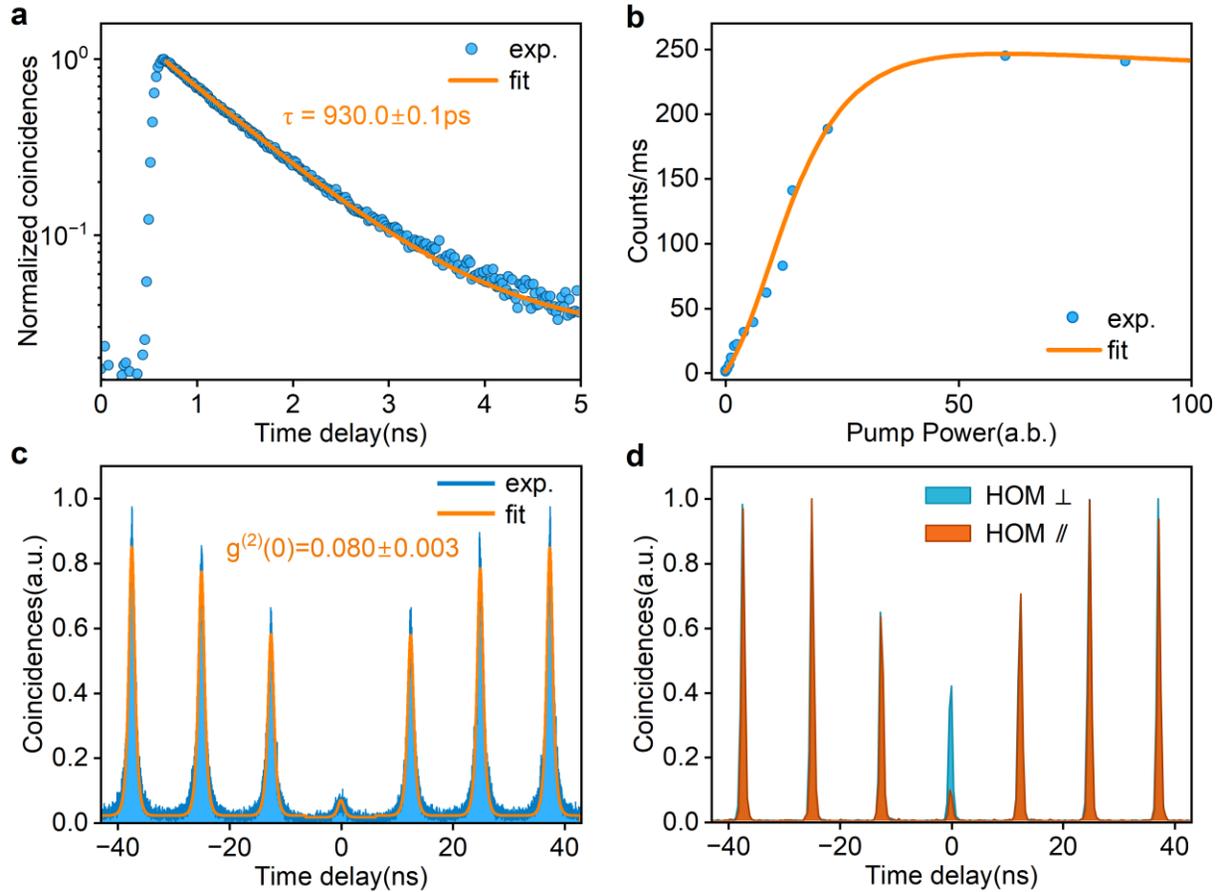

**Figure 4. Single-photon source characterization.** (a) Time-resolved photoluminescence decay curve, used to determine the radiative lifetime of the QD. (b) Power-dependent photoluminescence saturation curve, used to estimate the single-photon extraction efficiency. (c) Second-order photon autocorrelation histogram ($g^{(2)}(\tau)$) of the QD emission, showing strong antibunching at zero delay. (d) HOM interference: The visibility of 0.769(4) is extracted from second order correlation measurement of two consecutively emitted photons separated by ~12.5 ns.

## Discussion

We demonstrate heterogeneous integration of quantum emitters (InAs QDs) with low-loss LT photonics via micro-transfer printing. The butt-coupled waveguide architecture achieves >85% inter-waveguide mode transfer efficiency, enabling the first hybrid quantum photonic chip capable of cryogenic single-photon routing, a critical milestone for on-chip quantum information processing. Ultimately, full-scale quantum information processing demands ultra-low-loss delay lines (~1 dB/m) and integrated superconducting detectors (e.g., SNSPDs), both challenged by LT's current propagation loss and fabrication complexity (*45*).

Heterogeneous integration with SiN waveguides (*46*, *47*) or Ge-silica waveguides (*48*) and the transfer printing of monolithic SNSPDs (*49*) onto LN substrates provide viable solutions. Our micro-transfer printing technique inherently supports such multi-material stacking (*50*). As ultra-low-loss photonics and cryogenic detector technologies mature, this convergence enables fully integrated quantum processors for boson sampling, quantum walks, and entanglement distribution—unlocking optical quantum computing's potential within near-term experimental reach.

Beyond this initial demonstration, the hybrid integration of QD emitters with high-speed LTOI photonics opens several transformative avenues for scalable quantum information processing. First, the high-speed EO modulators enable the realization of demultiplexed single-photon sources. By dynamically routing temporally separated photons from a single QD into different spatial modes via electro-optic switches, one can efficiently generate the multi-photon states required for large-scale Boson sampling (*29, 51, 52*). Combined with low-loss optical delay lines to compensate for timing differences, this active demultiplexing approach offers a deterministic pathway to high-photon-number experiments. Second, this platform is well-suited for fusion-based quantum computing. The deterministic generation of entangled photon strings from QDs, coupled with on-chip linear optical circuits, allows for fusion measurements to link small-scale entangled states into large-scale cluster states (*53, 54*). This capability paves the way for measurement-based quantum computing by creating the necessary large-scale entangled resources on a chip. Furthermore, single-photon frequency shifting presents another critical perspective. By leveraging the strong second-order nonlinearity ($\chi^{(2)}$) of the LTOI/LNOI platform, visible or near-infrared photons emitted by QDs can be coherently converted to the telecom C-band (*55*). This frequency conversion is essential for interfacing high-performance on-chip quantum processors with long-distance fiber-optic quantum communication networks.

**Acknowledgments**

**Funding:**
National Natural Science Foundation of China (NSFC) (62435016, 62135012, 12374476)
The innovative research program (22-ZZCX-067, JS25-023)

**Author contributions:**
T.J., Y.C., L.L., and J.L. conceived the project. Y.C. and X.L. developed the methodology. K.X., D.S., and X.L. fabricated devices. X.L., K.X., and D.S. performed the measurements. Y.C., X.L., K.X., and D.S. analyzed the data. K.X., D.S., and X.L. contributed to visualization. P.C., W.W., Y.Y., J.W., and Z.W. supervised the project. Y.C., X.L., K.X., and D.S. wrote the original manuscript. All authors participated in reviewing and editing the manuscript. The paper has been polished with AI.

**Competing interests:** Authors declare that they have no competing interests.

**Data and materials availability:**
All data are available in the main text or the supplementary materials.